\documentclass[preprint]{aastex}
\usepackage{natbib}
\usepackage{url}
\usepackage{amsmath}
\usepackage{multirow}
\usepackage[symbol*]{footmisc}

\newcommand{\sat}[1]{\it\uppercase{#1}\rm}
\newcommand{\fig}[1]{Figure~\ref{#1}}
\newcommand{\tbl}[1]{Table~\ref{#1}}
\newcommand{\speed}[1]{#1 km s${}^{-1}$}
\newcommand{\aspeed}[1]{$\sim$#1 km s${}^{-1}$}

\shorttitle{Coronal Implosion and Particle Acceleration} %

\shortauthors{Liu et al.}

\begin{document}

\title{Coronal Implosion and Particle Acceleration in the Wake of a Filament Eruption}
\author{Rui Liu \& Haimin Wang}
\affil{Space Weather Research Laboratory, Center for Solar-Terrestrial Research, NJIT, Newark, NJ
07102; rui.liu@njit.edu}

\begin{abstract}
We study the evolution of a group of \sat{trace} 195 {\AA} coronal loops overlying a reverse
S-shaped filament on 2001 June 15. These loops were initially pushed upward with the filament
ascending and kinking slowly, but as soon as the filament rose explosively, they began to contract
at a speed of \aspeed{100}, and sustained for at least 12 min, presumably due to the reduced
magnetic pressure underneath with the filament escaping. Despite the contraction following the
expansion, the loops of interest remained largely intact during the filament eruption, rather than
formed via reconnection. These contracting loops naturally formed a shrinking trap, in which hot
electrons of several keV, in an order of magnitude estimation, can be accelerated to nonthermal
energies. A single hard X-ray burst, with no corresponding rise in \sat{goes} soft X-ray flux, was
recorded by the Hard X-ray Telescope (HXT) on board \textit{Yohkoh}, when the contracting loops
expectedly approached the post-flare arcade originating from the filament eruption. HXT images
reveal a coronal source distinctly above the top of the soft X-ray arcade by $\sim$15$''$. The
injecting electron population for the coronal source (thin target) is hardening by $\sim$1.5 powers
relative to the footpoint emission (thick target), which is consistent with electron trapping in
the weak diffusion limit. Although we can not rule out additional reconnection, observational
evidences suggest that the shrinking coronal trap may play a significant role in the observed
nonthermal hard X-ray emission during the flare decay phase.
\end{abstract}


\section{Introduction}
In the standard flare model \citep[CSHKP model;][]{carmichael64, sturrock66, hirayama74, kp76}, the
initial driver of the flare process is a rising filament above the polarity inversion line. The
rising filament stretches the overlying field lines, resulting in the formation of a current sheet
underneath and the consequent cancelation of magnetic fluxes of opposite polarity. It is
energetically impossible, however, to open up the closed magnetic field lines via a purely
ideal-MHD process in the force-free configuration \citep[the Aly-Sturrock paradox;][]{aly84,
sturrock91}. One way to avoid this constraint is to open only a portion of the closed field lines
\citep[see the review by][]{lsb03}. Considering a flux rope confined by an overlying magnetic
arcade, \citet{sturrock01} demonstrated that it is energetically favorable for part of the rope to
erupt into interplanetary space.

This \emph{partial eruption} concept has been further investigated both observationally
\citep[e.g.,][]{gilbert00, ghb01, gal07, lag07, liu08, tripathi09} and numerically
\citep[e.g.,][]{manchester04, fan05, gf06, bfh06, fg07}, especially in the context of eruptions
driven by the kink instability. In particular, \citet{fan05} demonstrated that the evolution of the
kink instability facilitates the loss of confinement of a flux rope. Specifically, the writhing
rotation changes the direction of the rope apex relative to the overlying arcade, which helps the
flux rope to ``rupture'' through the arcade field. \citet{gf06} extended Fan's simulation in time
to show that the flux rope bifurcates into two, with one part erupting and the other staying
behind. In the latter simulation, one can see that overlying loops are pushed upward and aside as
the flux rope kinks and expands, and that following the rupture of the arcade, those overlying
loops that have not reconnected with the rope field quickly contract back to the core field, which
is reasonable, considering the reduction of the magnetic pressure in the core field due to the rope
escaping. This is reminiscent of the ``implosion'' conjecture proposed by \citet{hudson00}, to
which \citet{lwa09} lent support in observation. In that case, however, the reduction of the
magnetic pressure is presumably due to the release of the free magnetic energy during the early
phase of the flare.

In this Letter, we present observations of the coronal loop contraction in the wake of a filament
eruption. The eruption occurred in the active region NOAA 9502 (S27E43) on 2001 June 15, associated
with a \sat{goes} class M6 flare starting at about 10:00 UT. By studying the photospheric helicity
transportation, \citet{rcz05} concluded that the reverse S-shaped filament erupted due to the
accumulation of magnetic helicity exceeding the kink instability threshold. Here we concentrate on
a late-phase hard X-ray (HXR) burst recorded by \textit{Yohkoh} (see the \textit{Yohkoh} flare
catalogue\footnote{\url{http://gedas22.stelab.nagoya-u.ac.jp/HXT/catalogue/index.html}; HXT event
number: 26710}), which occurred about 26 minutes later in the same active region, following the
loop contraction. We will analyze the observations in Section 2, discuss the results in Section 3,
and make concluding remarks in Section 4.

\section{Observation and Analysis}

\subsection{Coronal Implosion}

\begin{figure}\epsscale{0.9}
\plotone{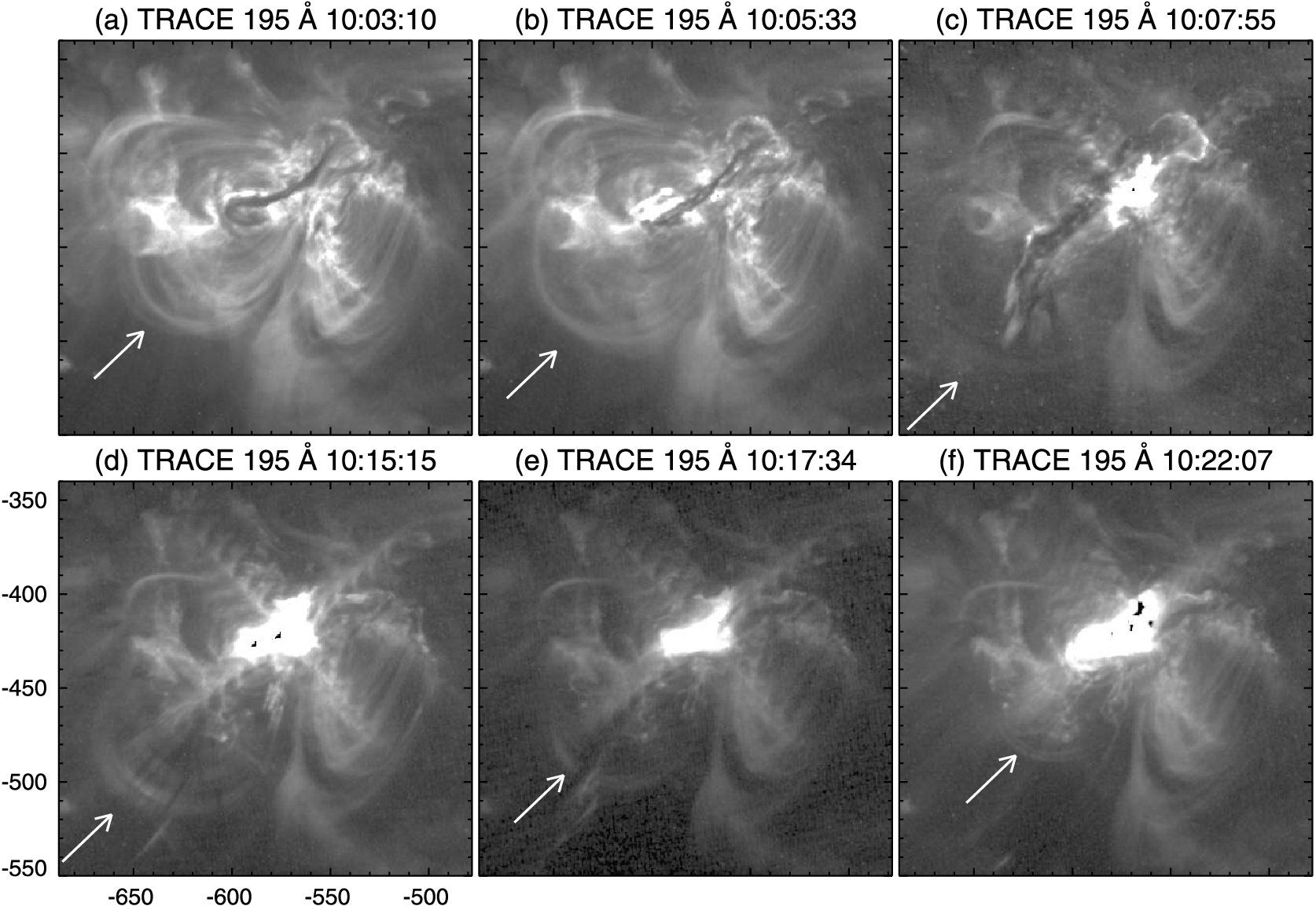} \caption{Snapshots of the expansion and contraction of coronal loops associated
with a filament eruption. Images are observed with the \sat{trace} 195 {\AA} filter. The loops of
interest are marked by white arrows. \label{trace}}
\end{figure}

\begin{figure} \epsscale{1}
\plotone{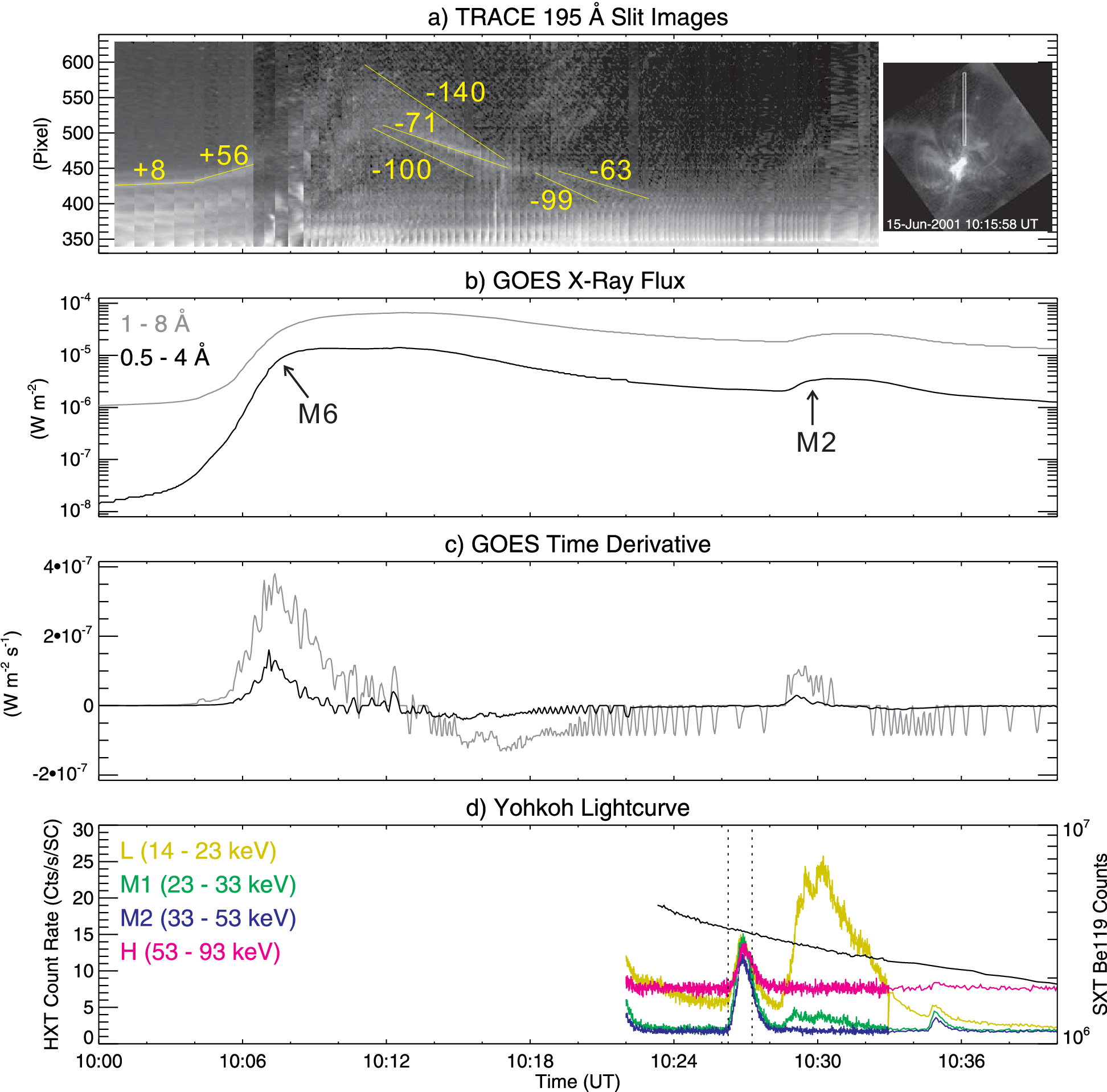} \caption{Coronal implosion observed in EUV and the lightcurves of the associated
flare. (a) Slices of \sat{trace} 195 {\AA} images cut by the slit as shown in the inset are placed
on the time axis. The \sat{trace} image in the inset has been rotated clockwise for $145^\circ$.
Labels indicate speeds derived from the apparent ``slopes'', in km s${}^{-1}$. (b) \sat{goes} soft
X-ray fluxes in 1--8 {\AA} (grey) and 0.5--4 {\AA} (dark). (c) Time derivatives of \sat{goes}
fluxes. (d) HXT counts per second and per subcollimator (SC) are scaled by the y-axis on the left,
and SXR counts recorded by the Soft X-ray Telescope (SXT) Be-119 filter (integrating over the
post-flare arcade of the M6 flare) are scaled by the y-axis on the right. Dotted lines mark the
time interval for the HXT image synthesis.\label{lcur}}
\end{figure}

The reverse S-shaped filament is obviously visible in \fig{trace}(a) as a dark, absorption
structure in the bright active region, as observed with the \sat{trace} 195 {\AA} filter. A bundle
of brightening loops astride the filament are marked by an arrow. The loops are supposed to connect
the opposite polarity separated by the filament channel (c.f. \S2.2). From about 09:57 UT to 10:06
UT, as the filament ascended slowly with a counterclockwise rotation of its axis \citep[also
see][]{green07}, the overlying loops are observed to rise with it (\fig{trace}(a--c)), presumably
being pushed upward by the filament, but to contract (\fig{trace}(d--f)) from 10:11 UT onward,
after the filament ``broke'' through the arcade at about 10:08 UT (\fig{trace}(c)); a comprehensive
animation of \sat{trace} images for this event is available at the \sat{trace}
website\footnote{\url{http://trace.lmsal.com/POD/bigmovies/filaments/T195_2001Jun15New.mov}}).
Lightcurves of the flare associated with the filament eruption are presented in \fig{lcur}. Within
one hour of the eruption, one can see that the filament channel in H$\alpha$ was filled with dark
material again (\fig{img}(d)), and that post-flare loops are shown as dark features perpendicular
to the channel, suggestive of the partial eruption nature of the event.

To study the loop evolution in detail, we rotate \sat{trace} images clockwise by $145^\circ$ so
that the loops of interest are positioned upward (see the inset of \fig{lcur}(a)). A slit of 290
pixels long and 10 pixels wide is placed across the group of coronal loops, and the slices of
\sat{trace} images cut by the slit are placed on the time axis in \fig{lcur}(a). Initially, the
loops expanded slowly at about \speed{8}, and then at a faster speed of about \speed{56}, roughly
the same speed of the rising filament. At about 10:06 UT, the filament started to rise upward
explosively, simultaneously with the onset of the flare impulsive phase, as indicated by the time
derivatives of \sat{goes} soft X-ray (SXR) fluxes (\fig{lcur}c), which usually resemble the HXR
profile \citep[Neupert effect;][]{neupert68}. At 10:07:55 UT (\fig{trace}(c)), the filament already
lost its coherent shape, and the slit was filled by dark filament material during 10:06--10:08 UT
(\fig{lcur}(a)). The group of loops became very diffuse and faint due to shorter exposures with
enhanced brightening in the flaring region. But one can still see that as early as about 10:11 UT,
the loops began to contract inward at various speeds ranging from \speed{60} to \speed{140}. As an
instructive comparison, the contraction of coronal loops observed in the flare early phase is at
\aspeed{5} \citep{lwa09}, while the shrinkage speed of relaxing cusp-shaped field lines observed in
SXRs is only \aspeed{2} \citep[][and references therein]{lin04}.

The contraction is observed to sustain for at least 12 minutes from 10:11 to 10:23 UT. During that
period the loops of interest contracted from the maximum projected height of $\sim1.3\times10^{10}$
cm, with respect to the flare footpoints (see \fig{img}(e)), to a height of $\sim6\times10^9$ cm,
where they became hardly visible, being overwhelmed by the bright active region. The post-flare
arcade originating from the M6 flare is about $4\times10^9$ cm high (projected). At an average
speed of \speed{100}, the contracting loops are expected to be stopped above the arcade in
$\geq200$ s, i.e., at about 10:26 UT, coincident with the HXR burst.

\subsection{Above-The-Loop-Top Hard X-Ray Source}

\begin{figure}\epsscale{0.82}
\plotone{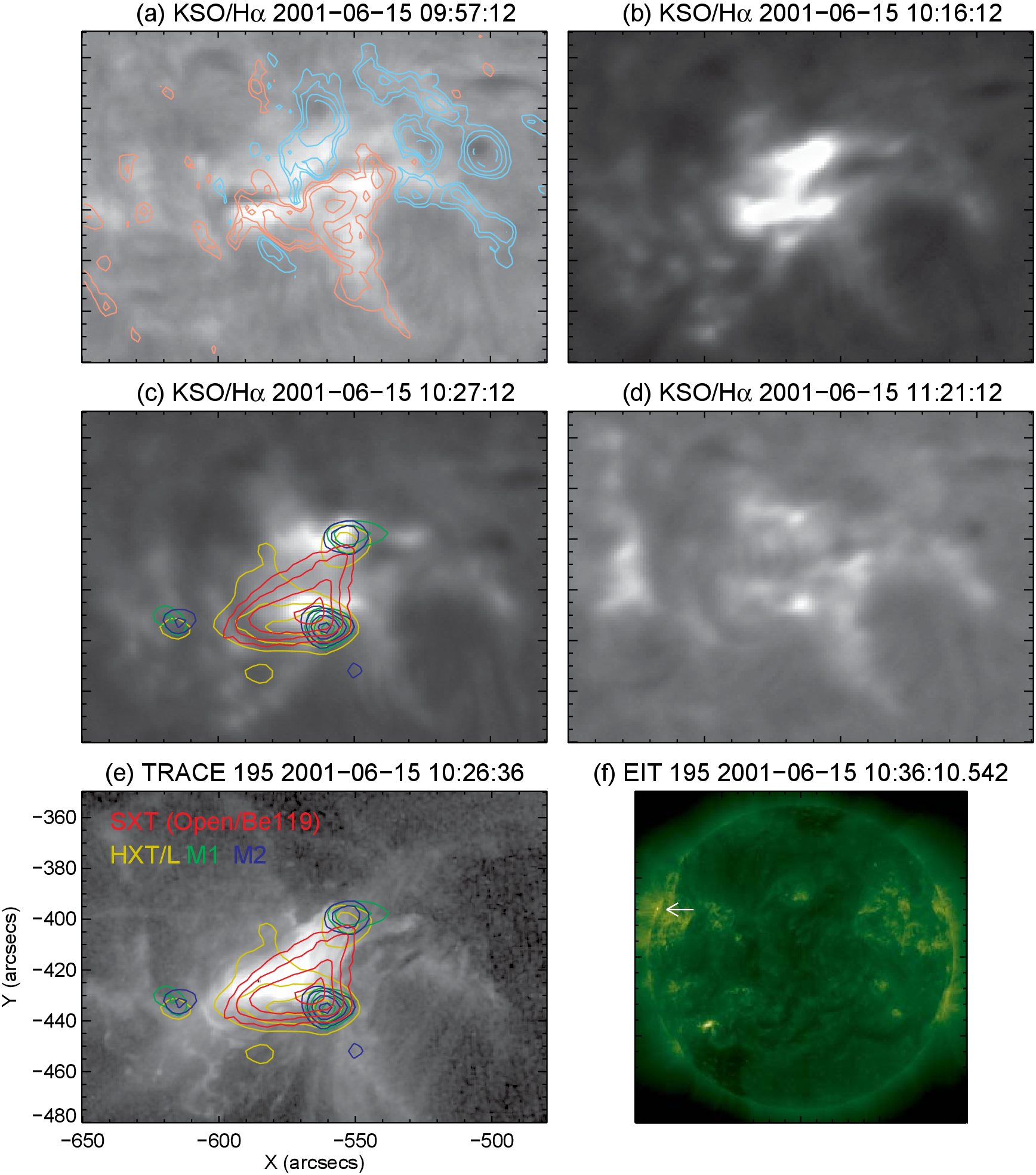} \caption{Multi-wavelength observations of the M6 flare and the late-phase HXR
burst. (a) A pre-flare H$\alpha$ image taken by the the Kanzelh\"{o}he Solar Observatory (KSO) is
overlaid with an line-of-sight magnetogram obtained by the Michelson Doppler Imager (MDI) on board
\sat{soho} at 09:35 UT. Contours levels are 100, 200, 400 and 800 G for positive polarities
(reddish orange), and -800, -400, -200, and -100 G for negative polarities (sky blue) (b) A
H$\alpha$ image shows the classical two-ribbon flare associated with the filament eruption. (c) HXT
and SXT contours for the late-phase burst are overlaid on the KSO H$\alpha$ image taken at
approximately the same time. The contour levels are 10, 20, 40 and 80\% of the maximum brightness.
(d) shows a post-flare H$\alpha$ image. (e) The same contours as in (c) are overlaid on a
\sat{trace} 195 {\AA} image taken at approximately the same time. The field of view is the same for
panels (a)--(e). (f) shows a full-disk \sat{soho} EIT 195 {\AA} image taken at 10:36 UT. A
brightening at the east limb is marked by an arrow. \label{img}}
\end{figure}

\begin{deluxetable}{ccccccccc}
\tablecolumns{9} %
\tabletypesize{\scriptsize} %
\tablewidth{0pt} %
\tablecaption{Spectral Characteristics of the late-phase HXR burst \label{spex}} %
\tablehead{\multirow{2}{*}{Algorithm} & \multicolumn{2}{c}{$\gamma^\mathrm{CS}$} &
\multicolumn{2}{c}{$\gamma^\mathrm{SFP}$} & \multicolumn{2}{c}{$\gamma^\mathrm{NFP}$} &
\multicolumn{2}{c}{T${}^\mathrm{CS}$ (MK)} \\ %
& \colhead{L/M1} & \colhead{M1/M2} & \colhead{L/M1} & \colhead{M1/M2}  & \colhead{L/M1} &
\colhead{M1/M2} & \colhead{L/M1} & \colhead{M1/M2}}%
\startdata%
MEM & \nodata & 3.7$\pm$1.1 & 1.6$\pm$0.4 & 3.2$\pm$0.3 & \nodata & 3.1$\pm$0.6 & \nodata & 163$\pm$64 \\%
Pixon & \nodata & 3.1$\pm$0.7 & 1.8$\pm$0.3 & 2.6$\pm$0.2 & \nodata & 2.3$\pm$0.3 & \nodata & 218$\pm$94 %
\enddata%
\tablecomments{Spectral information from HXT images is derived by calculating flux ratios between
adjacent energy bands. A region growing method is used to specify the integration region for each
individual HXR source \citep[see][for details]{lxw09}. CS stands for coronal source. SFP (NFP) is
referred to the footpoint source located in the south (north).}
\end{deluxetable}

\begin{figure} \epsscale{0.6}
\plotone{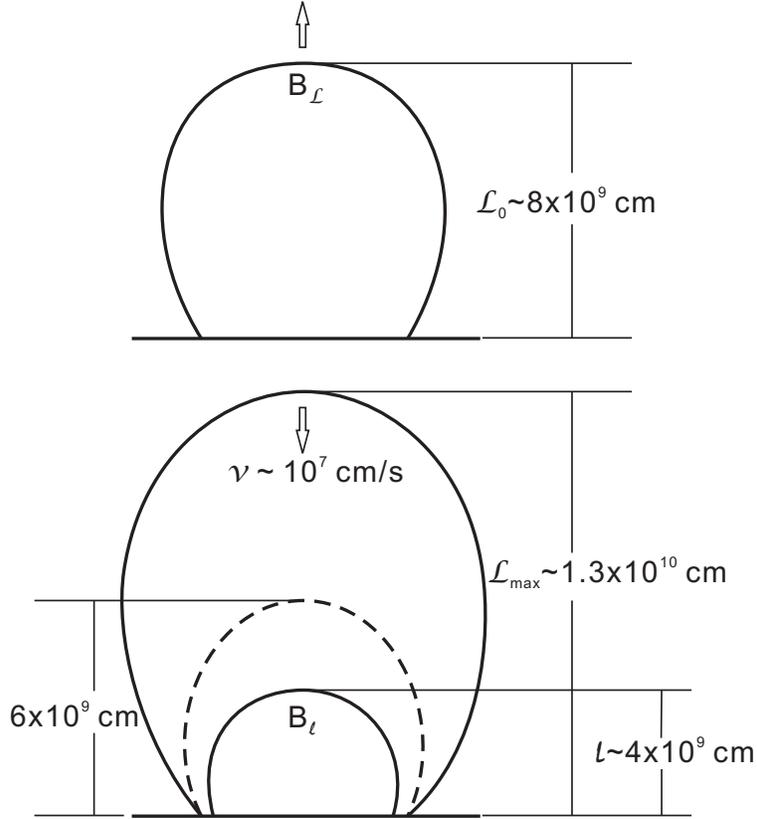} \caption{Cartoon of a shrinking trap as a simplification of the observations.
Top panel shows that initially the loop is pushed upward. Bottom panel shows that the loop shrinks
toward the post-flare arcade. $L_0$ is the loop's initial height ($\sim8\times10^9$ cm) measured
from \sat{trace} images. Similarly, $L_{max}$ is the loop's maximum height ($\sim1.3\times10^{10}$
cm), and $l$ the height of the post-flare arcade ($\sim4\times10^9$ cm). $v$ is the average
contraction speed. $B_l$ and $B_L$ denote the magnetic field at the height of $l$ and $L_0$,
respectively. The dotted line indicates the loop position at 10:23 UT.\label{cartoon}}
\end{figure}

The HXR burst observed by \textit{Yohkoh} HXT features a single peak that lasts for about 2 min.
The four HXT bands show a similar profile (\fig{lcur}(d)), with no corresponding rise in SXRs
(\fig{lcur}(b) and (c)). The dotted lines in \fig{lcur}(d) mark the time interval of data
accumulation for reconstructing HXR sources. The images synthesized with the Maximum Entropy Method
(MEM) are shown as contours in \fig{img} (c) and (e), which are very similar to those reconstructed
with the Pixon method (therefore are not shown here). Note three minutes later ($\sim$10:29 UT), an
M2 flare was recorded by \sat{goes}, which, however, is pinpointed to a different active region
(NOAA 9506), in view of the following observations: 1) The \sat{soho} EIT 195 {\AA} image taken at
10:36 UT (\fig{img}(f)) shows a brightening point at the east limb (marked by an arrow), with no
corresponding brightening in the previous EIT image at 10:24 UT (not shown); 2) HXT L- and M1-band
images (not shown) synthesized during the M2 flare also exhibit a point source co-spatial with the
195 {\AA} brightening; 3) The SXR counts integrating over the post-flare arcade of the M6 flare
decrease exponentially during the M2 flare (\fig{lcur}(d)).

The most striking feature of HXR burst is the existence of a coronal HXR source whose centroid
position is about $15''$ above the SXT loop top (\fig{img}(e)). No significant SXR emission was
detected at the coronal source location, indicating a rather low thermal plasma density. In
addition, the coronal source is so lacking in L-band emission that no plausible spectral index can
be derived from the L and M1 channel ratio (\tbl{spex}). In fact, the whole HXR burst is also
relatively lacking in L-band photons (see \fig{lcur}(d)). This indicates a short of low-energy
electrons, which normally would have constitute the bulk of the total energy. Thus, the energy
input into the chromosphere may be too small to drive chromospheric evaporation, which explains the
absence of the Neupert effect \citep{veronig05,warmuth09}.

On the other hand, the spectral index of the coronal emission derived from the M1 and M2 channel
ratio is rather hard, similar to those of the footpoints (\tbl{spex}). As a result, thermal
interpretation cannot yield plausible temperatures. One may wonder whether the presumed coronal
source is actually a third footpoint. However, this is not supported by the bipolar magnetic
configuration involved in this classical two-ribbon flare (\fig{img}(a)--(c)), nor by the apparent
bipolar coronal loops in the active region (\fig{trace}). Moreover, one can see in \fig{img}(c)
that the conjugate footpoints are located at the two flare ribbons, while no visible H$\alpha$
brightening is observed at the coronal source location.

If we assume the thin-target hypothesis for the coronal source
($\delta^\mathrm{CS}=\gamma_2^\mathrm{CS}-1$\footnote{We denote the photon spectral index derived
from the L- and M1-band ratio as $\gamma_1$, and that from the M1- and M2-band as $\gamma_2$, with
the superscripts, CS and FP, standing for the coronal source and the footpoint, respectively.
Corresponding electron power-law indices are denoted by $\delta$ in a usual fashion.}), and
thick-target for the footpoint emission ($\delta^\mathrm{FP}=\gamma_2^\mathrm{FP}+1$), then the
injecting electron population for the coronal source, $\delta^\mathrm{CS}$, is hardening by about
1.5 powers, relative to that for the footpoint, $\delta^\mathrm{FP}$, which is consistent with
electron trapping in the weak diffusion limit \citep[c.f., the Apendix in][]{ma99}.

\section{Discussion}

The characteristics of the above-the-loop-top source presented in this Letter are reminiscent of
the famous Masuda flare \citep{masuda94}, which is regarded as a very unique event \citep[see the
review by][]{krucker08}. In a recent revisit of the Masuda flare \citep{lxw09}, however, we noticed
that in the calibrated HXT images \citep{skm99}, the coronal source looks like a normal
\emph{loop-top} source, located slightly above the SXT loop top but within the 10\% brightness
contour of the HXT L-band thermal loop, rather than \emph{distinctly above the loop top} as in the
original HXT images \citep[e.g., Figure 3 in][]{aschwanden96}. Accordingly, its spectral indices
also fall into the normal range of coronal emission \citep[4--7;][]{tomczak01, kl08}. These changes
have relieved, in large measure, the longstanding difficulties of understanding the original Masuda
flare, viz., the lack of L-band emission of the coronal source, and the low density at the coronal
source location.

On the other hand, the so called above-the-loop-top sources reported in the \sat{rhessi} era have
been primarily, if not exclusively, found in a double coronal source morphology \citep{sh03, sui04,
pick05, veronig06, lg07, liuw08}. In that case, the lower coronal source is located at the thermal
loop top, while the upper source can be located as far as $30''$ above the lower one, moving upward
at a speed as high as 300 km s${}^{-1}$ \citep[e.g.,][]{sh03}. The two sources ``mirror'' each
other with respect to a presumed X-point reconnection site, in that higher energy emission comes
from lower altitudes for the upper source, while the lower source exhibits a reversed order
\citep[e.g.,][]{sh03,liuw08}. The spectra of the two sources have a similar power-law index
\citep{liuw08}, suggesting that the injecting populations are intimately related, presumably
accelerated at the X-point. If an X-point configuration were assumed below the above-the-loop-top
source studied here, as illustrated in \fig{img}(c), an additional loop-top source should also be
expected, which is obviously not the case.

Alternatively, the contracting loops may be relevant in the HXR production, since a shortening of
the field line length indicates a reduction of the magnetic energy intensity, which is available as
free energy to heat and accelerate particles. This effect has been discussed in the context of
newly reconnected cusp-shaped lines relaxing into potential-like lines at lower altitudes
\citep{sk97}.

As a simplification of our observation, \fig{cartoon} illustrates that a magnetic loop is initially
stretched and pushed upward (top panel), and later forms a shrinking trap (bottom panel) in which
mirroring particles are accelerated in a way similar to the trap-without-shock scenario in
\citet{sk97}. The lifetime of our shrinking trap,
\[
T\simeq\frac{L_{max}-l}{v}\simeq\frac{9\times10^9\ \mathrm{cm}}{10^7\ \mathrm{cm}\
\mathrm{s}^{-1}}=900 \quad \mathrm{s},
\]
where $v$, the shrinking speed of the coronal loop, is approximated to be \speed{100} (see
\fig{lcur}(a)). The collisional deflection time of an electron of energy $E$ is given by,
\[
t(E)=0.95\left(\frac{E}{1\ \mathrm{keV}}\right)^{3/2}\left(\frac{n_e}{10^8\
\mathrm{cm}^{-3}}\right)^{-1}\left(\frac{20}{\ln \Lambda}\right) \quad \mathrm{s},
\]
where the Coulomb logarithm, $\ln\Lambda=\ln(8\times10^6\,T_e\,n_e^{-1/2})\simeq20$ under typical
coronal conditions. Thus, for 10--100 keV electrons in the coronal trap with a density in the order
of $10^8$ cm${}^{-3}$, the trapping time is of order 100--1000 s, comparable to the trap lifetime.

For an 1--10 keV electron, its thermal speed $v_e\simeq(2-6)\times10^9$ cm s${}^{-1}$, so its
characteristic time between two subsequent reflections, $\tau\leq2L_{max}/v_e\simeq (4-13)$ s.
Since $\tau\ll T$, the conditions of the periodic longitudinal motions change adiabatically. The
longitudinal adiabatic invariant is therefore conserved \citep[][Chapter 6]{somov06}, i.e.,
\[
P_\parallel(t)\,L(t)=\mathrm{const}.
\]
Since $L_{max}/l\simeq3.3$, when $t\to T$, $P_\parallel(t)\to 3.3\,P_\parallel(0)$. In addition,
the transversal adiabatic invariant is conserved in the non-relativistic approximation,
\[
\frac{P_\perp^2(t)}{B}=\mathrm{const}.
\]
Assuming that both the initial and the final state of the magnetic loop are quasi-potential, we
estimate the field strength at the loop top, following the empirical formula from \citet{dm78},
i.e.,
\[
B_l\simeq0.5\,(l/R_\sun)^{-3/2}\simeq36 \quad \mathrm{G},
\]
and $B_L\simeq0.5\,(L_0/R_\sun)^{-3/2}\simeq13$ G, where $B_l$ and $B_L$ denote the magnetic field
at the height of $l$ and $L_0$, respectively. As the contracting loop approaches the post-flare
arcade (bottom panel of \fig{cartoon}), $B\to B_l$, hence $P_\perp(t)\to 1.7P_\perp(0)$, and
$E_k(t)=(P_\perp(t)^2+P_\parallel(t)^2)/2m_e \to 14E_k(0)$. Thus, hot electrons of several keV can
be accelerated to nonthermal energies in this trap.

\section{Conclusion}
We suggest that the escaping of the kinking filament in the 2001 June 15 event results in the
contraction of the overlying coronal loops, which can be regarded as a variant of coronal implosion
\citep{hudson00,lwa09}. The contracting loops form a natural shrinking trap, which can efficiently
accelerate electrons. This explains the HXR burst during the flare decay phase, which is
characterized by nonthermal coronal emission in consistency with electron trapping in the weak
diffusion limit. Although magnetic reconnection cannot be excluded in this picture, the shrinking
trap model seems to offer a more self-consistent interpretation.

We expect this acceleration process to occur generally in eruptions in which the arcade field is
only partially opened. Free energy builds up in the arcade when field lines are pushed upward by
the eruptive flux rope, since the magnetic energy associated with the loop,
\[
W=\frac{1}{8\pi}\int{\int{B^2\,ds}\,dl},
\]
where $s$ is the loop cross section, assuming to be constant, and $l$ the loop length. As the
stretched field relaxes in the wake of the rope escaping, trapped particles gain energy from the
increase in parallel momentum, due to shortening field lines, and from the increase in
perpendicular momentum, due to strengthening magnetic field at lower altitudes.

This process may play a role in some of the HXR bursts observed during the flare late phase, which
often show weak soft emission, hardening spectra, and/or high coronal sources
\citep[e.g.,][]{fd71,hudson78,gb08,warmuth09}. Such late-phase bursts have been often attributed to
the acceleration and trapping of electrons in the post-flare loop systems
\citep[e.g.,][]{cliver86}, or, to the acceleration of electrons in a shock front
\citep[e.g.,][]{fd71}. The mechanism suggested here, however, involves a transfer of the free
magnetic energy from the core field to the confining arcade during the eruptive process. The energy
transferred is then made available to the trapped particles in the aftermath of the partial
eruption, through the contraction of the loops that have not reconnected.

\acknowledgments The authors thank the anonymous referee for helpful comments. We would like to
acknowledge the \sat{trace} and \textit{Yohkoh} consortia for the excellent data. RL thanks J.~Lin
for helpful discussion, and Y.~Xu for processing KSO H$\alpha$ data, which are provided through the
Global High Resolution H-alpha Network. This work was supported by NASA grant NNX08-AJ23G and
NNX08-AQ90G, and by NSF grant ATM-0849453.


\end{document}